\documentclass[sigconf]{acmart}
\usepackage{cleveref}

\setcopyright{cc}
\copyrightyear{2025}
\acmYear{2025}
\setcctype{by}

\acmConference[ITiCSE 2025]{Proceedings of the 30th ACM Conference on
Innovation and Technology in Computer Science Education V. 1}{June 27-July 2,
2025}{Nijmegen, Netherlands}

\acmBooktitle{Proceedings of the 30th ACM Conference on Innovation and
Technology in Computer Science Education V. 1 (ITiCSE 2025), June 27-July 2,
2025, Nijmegen, Netherlands}

\acmDOI{10.1145/3724363.3729090}

\acmISBN{979-8-4007-1567-9/2025/06}

\newcommand\UniversityofGlasgow{University of Glasgow}
\newcommand\SchoolofComputingScience{School of Computing Science}
\newcommand\JPMorgan{JP Morgan Chase Technology}

\begin{document}

\title[Perspectives of Industry Practitioners as Educators]{%
  Out of the day job: Perspectives of Industry Practitioners in Co-Design and Delivery of Software Engineering Courses}


\author{Gillian Daniel}
\email{gillian.a.daniel@jpmorgan.com}
\affiliation{%
  \institution{JP Morgan Chase Technology}
  \city{Glasgow}
  \country{United Kingdom}
}

\author{Chris Hall}
\email{chris.w.hall@jpmorgan.com}
\affiliation{%
  \institution{JP Morgan Chase and Co}
  \city{Glasgow}
  \country{United Kingdom}
}

\author{Per Hammer}
\email{per.hammer@jpmchase.com}
\affiliation{%
  \institution{JP Morgan Chase and Co}
  \city{Glasgow}
  \country{United Kingdom}
}

\author{Alec-Angus Macdonald}
\email{alec-angus.macdonald@jpmorgan.com}
\affiliation{%
  \institution{JP Morgan Chase Technology}
  \city{Glasgow}
  \country{United Kingdom}
}

\author{Hollie Marwick-Best}
\email{hollie.c.marwick-best@jpmchase.com}
\affiliation{%
  \institution{JP Morgan Chase Technology}
  \city{Glasgow}
  \country{United Kingdom}
}

\author{Emma Mckenzie}
\email{emma.mckenzie@jpmchase.com}
\affiliation{%
  \institution{JP Morgan Chase and Co}
  \city{Glasgow}
  \country{United Kingdom}
}

\author{George Popa}
\email{george.popa@jpmorgan.com}
\affiliation{%
  \institution{JP Morgan Chase and Co}
  \city{Glasgow}
  \country{United Kingdom}
}

\author{Derek Somerville}
\email{derek.somerville@glasgow.ac.uk}
\affiliation{
  \institution{University of Glasgow}
  \city{Glasgow}
  \country{United Kingdom}
}
\author{Tim Storer}
\email{timothy.storer@glasgow.ac.uk}
\affiliation{
  \institution{University of Glasgow}
  \city{Glasgow}
  \country{United Kingdom}
}

\begin{abstract}

  Over more than two decades, The \UniversityofGlasgow~ has co-designed and delivered numerous software engineering
  focused courses with industry partners, covering both technical and discipline specific professional
  skills.
  Such collaborations are not unique and many of the benefits are well recognised in the literature. These include
  enhancing the real-world relevance of curricula, developing student professional networks ahead of graduation and
  easing recruitment opportunities for employers.

  However, there is relatively little scholarship on the perspectives
  of industry practitioners who participate in course design and delivery.  This gap is significant, since the effort
  invested by practitioners is often substantial and may require ongoing support from both the industry partner and
  academic institution. Understanding the motivations, expectations and experiences of practitioners who engage in
  course delivery can guide the formation of future partnerships and ensure their long-term sustainability.
  
  We begin to address this gap by reporting on the outcomes of a retrospective conducted amongst the practitioner
  coauthors of this paper, with the academic coauthors acting as facilitators. All coauthors have participated in the
  recent co-design and delivery of software engineering courses, but we choose to focus explicitly on the perspectives
  of the practitioners.  We report on the themes that emerged from the discussions and our resulting recommendations for
  future collaborations.

\end{abstract}




\maketitle

\renewcommand{\shortauthors}{Gillian Daniel et al.}


\section{Introduction}
\label{sec:introduction}


The \SchoolofComputingScience~ at The \UniversityofGlasgow~ has developed a broad programme that engages industry
partners in student education, both within curricula and wider extra-curricula activities. In particular for the focus
of this paper, we have at least two decades of experience of co-design of software engineering focused courses. These
courses have covered both technical and discipline specific professional skills and have been developed with a variety
of industry partners.  In our institution, learning objectives are typically agreed jointly, with academics then taking
responsibility for assessment design, whilst industry partners focus on the development and delivery of course content
such as lectures, laboratories and seminars.

The research literature has identified a number of benefits of co-designed courses for both academia and industry
\cite{samuel_universityindustry_2018, borah_are_2019, manisha_industry_2011}.  Most immediately, industry input is a
means for the industry partner to influence curricula and ensure that content is relevant to their specific needs.
However, co-design \cite{Kelly} is a means of balancing these needs with input from academia, to ensure that content is
relevant more broadly, whilst also ensuring that assessments are in accordance with relevant institutional standards.
Participation in co-design is also a way for the industry partner to demonstrate their commitment to corporate social
responsibility by ``giving back''. In some cases, having a presence on campus can also be a useful opportunity to
undertake recruitment \cite{Carless}.  Given the recognised benefits, industry partners may agree to second some of
their staff to the academic institution on a part time basis, in some cases making allowances in their main roles as
software professionals to allow them to dedicate sufficient time to the activity.

For an academic institution, co-design can help to ensure that a course is relevant to industry, since content is informed
by recent practice.  The existence of collaborations for the development of partnerships can be a means of
signalling to prospective students that degree programmes have industry relevance that can enhance future career
prospects.  In turn, this can enhance the employability of current students who can refer to industry relevant
skills on resum\'es and during interviews.  Students also benefit from gaining access to active industry practitioners,
enabling them to establish or broaden their nascent professional networks.

Despite the well-recognised benefits for both academic institutions and industry partners \citep{chen2016, Narayana}, the
experiences and perspectives of individual practitioners have not been addressed in the literature.  This gap is
significant, since industry practitioners take on considerable responsibility in designing and delivering a higher
education course, whilst still retaining most if not all of their responsibilities to their employer.  The success of a
co-design collaboration therefore depends significantly on the commitment, motivations and expectations of the industry
practitioners volunteering or seconded to the activity. Further, the long term success and sustainability of a
collaboration will be influenced by their experiences (both positive and negative) during course delivery. 
This includes both how they are supported by their employer and the support provided by the academic institution
in adjusting to the role of an educator.

\emph{Contribution:} %
The authors have collectively designed and delivered several courses at the \UniversityofGlasgow~ in recent years.  The
courses were co-designed with the same software industry partner - \JPMorgan. All of the authors have at least two years
experience of course co-design and delivery and in some cases many more.  To begin to address the lack of literature
from the perspective of industry practitioners, we present the results of a recent retrospective we conducted on the
topic among the coauthors.  The industry practitioners acted as participants in the retrospective, while the
academics took the role of facilitators, with the aim of ensuring that the conversation focused on industry rather than
academic perspectives.  We stress that the findings reported are the perspectives of the industry practitioners, and
should not be considered those of their employer organisation.

The rest of this paper is structured as follows.  \Cref{sec:related} summarises the previous research that has studied
the experiences, challenges and benefits of curricula co-design for different stakeholders, with particular reference
where possible to Computing Science, Software Engineering and closely related disciplines.  \Cref{sec:background}
sketches the context and approach to co-design undertaken by the coauthors, as well as outlining the two resulting
courses.  \Cref{sec:method} describes the method adopted for eliciting and analysing the perspectives of the
practitioner coauthors and \Cref{sec:results} presents the results of the activity.  Finally, \Cref{sec:conclusion}
draws lessons from the retrospective exercise for future co-design collaborations and presents our conclusions.


\section{Related Work}
\label{sec:related}


%

Many higher education institutions seek to involve industry partners in the institution's community and associated
educational activities. In Computing Science, Software Engineering and related disciplines, this may take place outside
of the strict curricula, such as participating as audience members in presentation, poster sessions or graduation prizes
or providing informal advice on curricula through review boards or similar, or acting as informal mentors or customers
for university based projects. In addition, industry partners may become more directly involved, through provision of
internship opportunities \citep{jaime_effect_2020} that are assessed within the curriculum structure, employment of
degree apprenticeship students \citep{smith_degree_2021} on work-based learning degrees or delivery of guest lectures
\cite{krogstie_guest_2018} or sponsorship of hackathon challenges
\cite{medina_angarita_what_2020,nolte_you_2018,Starov_uni_ind}.

\citet{Starov_uni_ind} highlighted hackathons helped generate ``industrial recruiting or generating ideas for
start-ups''. \citet{setubal_investigating_2024} highlighted industry must ``include opening opportunities
for beginners, such as trainees and interns''. Employability \citep{setubal_investigating_2024,%
  berbegal-mirabent_fostering_2020,garousi_challenges_2016,lucietto_academic_2021,manisha_industry_2011} had a common
theme that education institutions feel they needed to address by creating a stronger industry partnership. \citet{manisha_industry_2011} noted the ``exponential growth of employment opportunities'', and the IT industry
``felt a pressing need'' to acquire competent talent. Industry and education institution partnership
\citep{manisha_industry_2011} is required to enhance employability.

In some cases, and the focus of this paper, industry practitioners may be involved in the co-design and delivery of full
courses within a degree programme. \citet{zhuang_ushering_2024} discuss the strategy of Teaching-focused
University-Industry Collaborations (TFUIC) introduced in China. TFUIC aims to increase the partnership between
educational institution and industry. The strategy included ``joint creation and delivery of academic courses''
\cite{zhuang_ushering_2024}. This was felt to ``provide students with real-world exposure to complex industrial projects
but also enable companies to recruit well-prepared graduates''. It was felt this partnership helps keep courses
``contemporary and relevant'' \citep{zhuang_ushering_2024}. This was felt to be exemplified through the joint development
of the massive open online course (MOOC) titled Electronics Instrument Usage''.

The UK Department of Business, Innovation and Skills (BIS) \cite{bis_2010} created a report demonstrating the value of
postgraduates to the UK economy. It felt ``HEIs [Higher Education Institutions] need to be more pro-active in providing
postgraduates with the opportunity to develop the core competencies they need to succeed in a competitive job market''
and ``transferable skills training is embedded as standard in the funding and design of all postgraduate research
programmes'' \citep{bis_2010}.

\citet{samuel_universityindustry_2018} discuss the ``MSc in Structural Integrity'' created in partnership by Brunel
University London and The Welding Institute (TWI). The aim is ``to supply `work-ready' graduates" with technical and
professional skills. \citet{manisha_industry_2011} also highlight the recognition of ``soft skills'' or
professional skills in industry. By mid-course, 94\% \citep{samuel_universityindustry_2018} ``felt that their understanding
of up-to-date technology and industrial standards was beneficial to career development.'' Post course this dropped to
53\%, although this was felt to be recognised due to a ``sharp downturn in oil and gas industry recruitment``. Post course
75\% ``recorded having a good experience''.

\citet{borah_are_2019} discuss a degree with partnership between an educational institution and industry in India. It
specialised in the skills for research and development (R\&D) in Information and Communication Technology
(ICT). Graduates felt they required ``theoretical knowledge and (ICT) industry-specific practical and applied
skills''. They focussed on course based and project based collaboration. For the co-development there was "a series of
negotiations on the type and level of knowledge to be imparted to the students''. First the ``courses are developed
through face-to-face discussions between faculty and the firm's R\&D employees''. Second the developed courses are
reviewed by the "Board of Studies". Courses and projects were also designed entirely by the partner firm.

\citet{borah_are_2019} discuss co-delivery jointly by the partner firm and the educational institution. ``Faculty are
first trained by the partner for R\&D employees on the relevant topics, tools and teaching methodologies.''  The faculty
and firm share the responsibility for conducting assessment, the ``evaluation is completed'' by the faculty. The firm led
delivery from experts in specific technology domains were delivered through ``2-3 days of intensive training''. This was
to ``reduce the employees' time invested in organising weekly visits''.

\citet{manisha_industry_2011} reported that only ``25\% of Indian graduates are employable.'' It was felt the following
skills were lacking: technical skills, soft skills, process awareness and English proficiency. The partnership with
industry had the following expectations: self-learning, problem solving, ability to work in teams and diverse
perspectives to work. The company worked with took two approaches ``bottom up'' with faculty and students and ``top down''
working with governing bodies.


\section{Background}
\label{sec:background}


The authors of this paper have variously worked collaboratively on the co-design and delivery of educational activities
over a period of more than 15 years, for students within the \UniversityofGlasgow.  This initially comprised
extra-curricula activities such as hackathons and training activities such as small group resum\'e/CV review.  We have
been working together to deliver collaboratively designed courses within the curricula for approximately 10 years, with
the contents and focus under-going considerable evolution over this period.  Some of these collaborations have received
specific funding from the organisation (e.g. sponsorship of Hackathon challenges) whilst in other cases, we have secured
Visiting Professorships, funded by the Royal Academy of Engineering \citep{raeng} in the UK to support the development of course
materials.

An initial course was designed focusing on development of software for financial systems, reflecting the domain of
expertise of the industry provider.  However, in reviewing the course, a decision was taken to focus more closely on
enhancing student technical and discipline specific professional skills.  This led to the design of an advanced software
engineering course, with the intention being to cover content and material relevant to the contemporary software
industry.  The intention was that some of this content would gradually be transferred to earlier courses in the
\UniversityofGlasgow's degree programme and be replaced by new material as need arose.  Aspects of the courses have
previously been described by \cite{simpson2017experimenting-cseet}.

A final evolution took place when a decision was made to separate and substantially develop the technical, product release focused content and
professional skills material into two separate courses. Both courses rely extensively on laboratory based practice and
interactive seminars. In addition assessment is focused around coursework rather than exams. One, \emph{Software Product
  Release Engineering} (SPRE), covering DevOps practices, based in part on \citeauthor{Nygard2018ReleaseIt}'s
\citep{Nygard2018ReleaseIt} book.  This course runs during a single semester, with an approximate class size of 100
students.  Students are assessed on their ability to deploy and maintain a simple web application, whilst maintaining
high availability even as requirements evolve.  The course was initially created through the Visiting Professorship
scheme \citep{raeng}.  Consequently, the course is primarily delivered and assessed by an industry practitioner (Hammer)
with support from an academic course coordinator (Storer).

The second, \emph{Coaching Software Teams} (CST) focuses on the professional skills needed to foster software team
cohesion and performance.  Taught material is delivered in semester 1 by industry practitioners, and covers material
such as gaining legitimacy as a coach, managing retrospectives and coaching technical skills, such as pair
programming. The students participate as coaches to teams of junior students in our Team Project course (previously
described by \citet{simpson2017experimenting-cseet}) through the whole of the academic year. The student coaches are
assessed on their ability to coach process improvements within their team, with assessment administered by the academic
course coordinator.  There are approximately 40-60 students on this course each year, with coaches working individually
or in pairs with teams of junior students.

Throughout this history, the coauthors have had the support of their respective organisations in undertaking the
collaboration.  However, neither party has deemed it necessary to formalise the arrangement in a written agreement,
since there is a great deal of good will generated between the organisations.  Further, we'd also note that although the
authors have no principled objection to the use of co-delivered courses for promoting recruitment opportunities, this is
not something we have done at \UniversityofGlasgow. Students are free to approach the practitioners about opportunities
during course delivery if they wish, but these are not advertised explicitly.  In addition, other activities are
organised during the academic year, such as recruitment fairs and hackathons that provide explicit opportunities for
discussion about recruitment and careers.


\section{Method}
\label{sec:method}


To elicit the perspectives of our practitioner coauthors, we adopted a methodology similar to Agile Retrospectives
\citet{schwaber01agile}, using a structured process to gather data and then collectively analysing the issues
identified. Four authors (Hall, Hammer, Somerville, Storer) met to develop an initial set of questions for participants
in the retrospective.  Questions were brain-stormed and written and edited on a whiteboard.  Each question was then
added as a slide to a presentation deck.  We then circulated the deck of slides to a wider group asking them to review
the questions and add further suggestions, or propose amalgamations as desired.

Once the final set of questions was agreed on (see \Cref{fig:initial-questions}), the authors scheduled a retrospective at
the practitioners' premises to maximize attendance. Seven authors met (Hall, Hammer, Macdonald, Mckenzie, Popa,
Somerville and Storer) to conduct the retrospective.  The meeting lasted approximately 90 minutes, with Somerville and
Storer acting as facilitators.  The question deck was printed out on A4 sheets (one question per sheet) and participants
were invited to annotate the questions with responses, using sticky notes in order to gather initial data.  They were also encouraged to add different
questions (and answer them) if these occurred to them. Once the responses were complete, the sticky notes were grouped
into themes as chosen by the participants.

\begin{figure}
  \centering
  \fbox{
    \begin{minipage}{\linewidth}
      \raggedright
      \vspace{.2em}
      \begin{list}{-}{\setlength{\leftmargin}{1em}\setlength{\rightmargin}{1.2em}}
      \item How did you become involved in the course?
      \item What surprised you during the course?
      \item What did you learn from participating in the course?
      \item What practical challenges did you face?
      \item Has your expectation of students and/or graduates changed?
      \item Were there any benefits to your career from participating?
      \item What reasons would you give to convince someone else to participate in university education?
      \item What could you see yourself doing next in relation to education?
      \item What challenges were then in coordinating the delivery of the course?
      \end{list}
      \vspace{.2em}
    \end{minipage}
  }
    
  \caption{Initial questions to practitioners to elicit perspectives on participating in co-design of a
    software engineering focused course.}
  \label{fig:initial-questions}
  \Description[The initial questions prepared for data gathering.]{%
    The initial questions prepared for data gathering:
    How did you become involved in the course?; What surprised you during the course?; %
    What did you learn from participating in the course?; %
    What practical challenges did you face? Has your expectation of students and/or graduates changed?; %
    Were there any benefits to your career from participating?; %
    What reasons would you give to convince someone else to participate in university education?; %
    What could you see yourself doing next in relation to education?; %
    What challenges were then in coordinating the delivery of the course?}
\end{figure}

The participants then selected themes for further discussion collectively, which lasted for approximately 1 hour.  The
conversation was recorded for later review. The documentation generated was then recirculated, allowing the practitioner
coauthors to add further comment as desired.  The findings reported in the next section reflect both the initial written
comments, the transcript of the conversation, the further notes provided and the facilitators notes made during the
meeting.  The results presented below therefore represent the \emph{themes} discovered during the retrospective process,
rather than the original questions asked. Discussions focused on elaborating the meaning of the themes and identifying
actions that then constituted our recommendations.


\section{Findings}
\label{sec:results}


The practitioners reported that in general there was an eagerness to volunteer in educational activities amongst their
colleagues.  Motivations included ``giving back'' by providing benefit to students through both the formal education and
helping them to establish their professional networks. The practitioners also reported personal benefits, including 
learning something new, (``the best way to learn is to teach''),
the enhanced reputational benefits (both professionally and personally) of being formally registered as an institutional
affiliate and the practical benefits of being able to access University facilities.  The practitioners also noted the
collaboration as an opportunity to help ``shape'' the kinds of graduate engineers they themselves would like to work
with in the software industry in the near future.  This was distinct from the organisational desire to prepare graduates
for their specific recruitment needs, since the practitioners were motivated to better prepare graduates for the
software industry as a whole.  Overall, this meant the practitioner who coordinated the collaboration was generally not
short of volunteers to become involved in the course.

However, the practitioners also noted that their colleagues often underestimated the time commitment involved in
participation. For example, the CST course comprised a one hour lecture and a two hour interactive seminar each week,
which often involved multiple practitioners to implement effectively. Many colleagues assumed the commitment was limited
to the delivery of content, however, further effort was spent on other activities, such as iterative content
development, coordination, and review of assessments with the academics. This was exacerbated by the need to sometimes
redevelop material as elements of the co-designed courses began to appear in other courses in the University.  The
course coordinator therefore had to balance the desire to extend opportunities for participation in the courses with the
need to ensure volunteers were suitably committed, as well as the need to mentor and support new members of the team
during content development.

The practitioners reported several aspects of the experience as an educator as being beneficial to their day job.  On a
practical level, delivery of lecture content to students enabled practitioners to develop and practice their
presentation skills.  Reviewing the theoretical material to be incorporated in course content helped practitioners to
assess and deepen their own knowledge of the different topics and led them to reflect on the practices adopted
within their teams.  This could happen when preparing their own lecture content, or when watching a lecture that had
been prepared by a colleague.  In addition, practitioners would also prototype ideas for exercises with other groups and
teams, creating opportunities for them to experience how theoretical descriptions of practices were implemented
elsewhere in the organisation.

Practitioners reported that this exposure led them to consider how practices were interpreted and adopted differently
throughout the organisation.  Going further, they were stimulated to reflect on \emph{why} particular theoretical practices
weren't always adopted according to the theoretically ``right'' way internally. This was sometimes due to unavoidable
practical constraints, such as the nature of a particular project or resource availability. However, in other
cases reviewing theoretical materials enabled the practitioners to strengthen their advocacy for improved software
processes internally, by referring to specific theoretical descriptions of the desired approach.  As one practitioner
said,

\begin{quotation}

  ``usually someone said it better than I can explain it...I find it really useful having recently touched on
  refactoring and touching a Martin Fowler book. Now I was able to throw that back when I was trying to convince
  [colleagues].''
  
\end{quotation}

Some of the material developed in the course was also used internally within the organisation for training purposes.

The practitioners described other ways that participation in a course had been beneficial.  The team who delivered the
courses received the Employee Appreciation Award for the work they had done, indicating significant recognition within
the organisation.  Another practitioner was mentioned in an internal newsletter in connection with a course.  The
value of the collaboration was recognised in other ways, for example, other colleagues had taken the structure and
materials developed at the \UniversityofGlasgow~ and redeployed them at two other higher education
institutions. Participation in the course also created opportunities for the practitioners to meet with senior
management in order to gain support for the collaboration.  This created opportunities for practitioners to increase the
visibility of their contribution to the organisation, as well as extend their internal networks.

Practical challenges to course delivery were encountered by the participants.  For example, some of the IT
infrastructure provided by the institution was quite dated, leading to incompatibilities between equipment, such as
laptops and projectors.  Similarly, spaces allocated were not suitable for the style of teaching anticipated.  The
coauthors had focused on developing interactive seminars that required small-group working.  However, rooms allocated
often had fixed lecture-style layouts making discussions and collaboration difficult.

Participants noted difficulties with remembering passwords to institutional systems and generally gaining access to IT
resources such as wireless networks and learning resources.  Similarly, due to the nature of their organisation's business,
transferring artifacts between the organisation and the institution, such as lecture slide decks, was
difficult. These challenges perhaps stem primarily from the need to work in both the existing environment with the
associated organisational constraints and to intermittently bridge across from this into the academic institution.

Participants noted the unfamiliarity of bureaucratic processes and the complexity of reporting, ownership and
accountability of the co-designed course.  The arrangement of the collaboration between the academic partner and the
institution has historically been relatively informal.  The local office of the organisation has a long history of
positive engagement with the institution, with individual activities generally obtaining informal approval from relevant
line managers.  Although this provided flexibility, the approach can make it more difficult for practitioners to
navigate, since it is unclear who necessarily ``owns'' or ``sponsors'' activities internally.  This can make, for
example, seeking approval for time away from a practitioner's day job more difficult to secure, being dependent on the
priorities of line managers.  Although most were reported to be very supportive, this can also create inequality of
opportunity to participate in the courses.

Several participants reported their perceptions of the student cohorts that they had taught over successive sessions.
Most immediately, interaction with all the cohorts enabled the practitioners to recognise the value of their own
knowledge and experience.  As one of the practitioners said,

\begin{quotation}
  ``So you thought they know all the latest, but then when you started actually interacting with them, you realised they
  knew very little... and then I give my lecture and they had some questions. This was actually useful for them
  because the questions they put means that they're digging right now into this material.''
\end{quotation}

Exposure to student cohorts therefore enables practitioners to recognise the value of their own expertise and
experiences and calibrate this against incoming graduates.

Differences were noted between the different cohorts taught.  In 2023-24, on the CST course, most students were
graduate apprentices, who spend the majority of their degree programme learning in the workplace. For administrative
reasons, most students on the 2024-25 course were from the on-campus programmes.  Practitioners agreed that the graduate
apprentices generally had better developed professional skills than the on-campus cohort and found it easier to engage
with the practitioners.  In addition, the graduate apprentices tended to find the material presented easier to relate
to, because they had more real world experience of the issues addressed.  In addition, the practitioners found that
on-campus students tended to take a more transactional approach to their learning focused around their assessments.  For
example, the courses use an assessed in-person quiz to incentivize attendance at seminars. In the 2024 session, it was
noted that students would attend the seminar for the quiz and then immediately leave, whereas the 2023 (graduate
apprentice) cohort did not do this, perhaps due to a reluctance to participate in what might be perceived as rude or
unprofessional behaviour.


\section{Conclusion and Recommendations}
\label{sec:conclusion}


This paper has reported on the authors' collective reflection on the experience and perspectives of industry
practitioners in co-designing and delivering software engineering courses in higher education.  To the best of our
knowledge, we make a novel contribution through our focus on the perspective of the practitioners in this context,
rather than the experience of students or the benefits to the respective organisation, as can be found elsewhere in the
literature.

Our findings on the experiences also allow us to make a number of recommendations for the development of future
collaborations.  Our recommendations are summarised in Figure \ref{fig:recommendations}:

\begin{figure}
  \centering
  \fbox{
    \begin{minipage}{\linewidth}
      \raggedright
      \vspace{.2em}
      \begin{list}{-}{\setlength{\leftmargin}{1em}\setlength{\rightmargin}{1.2em}}
      \item Give consideration to the time commitment required for participation and accommodate availability in
        programme design.
      \item Recognise the benefits to both the practitioner and organisation of practitioner participation in course
        design and delivery.
      \item Determine what level of formal agreement is required between the institution and organisation to enable
        support and access to resources.
      \item Institutions should develop processes to fully involve practitioners as faculty wherever possible.
      \end{list}
      \vspace{.2em}
    \end{minipage}
  }
    
  \caption{Our recommendations for the future co-design and delivery of higher education courses with industry
    practitioners.}
  \label{fig:recommendations}
  \Description[Summary of recommendations that are fully described in the text.]{%
    Summary of recommendations that are fully described in the text.}
\end{figure}

\begin{itemize}
\item We noted that in general, industry practitioners are keen to participate in education but may underestimate the
  time commitment. For example, at least some of our course development has been supported by the Royal
  Academy of Engineering's Visiting Professorship scheme, which recognises the time commitment involved.  Consideration
  therefore needs to be given to mechanisms that can better spread load and opportunity across an organisation,
  potentially over multiple years.  In addition, the time commitment of course leaders (both industry and academic) in training and mentoring
  contributors to courses needs to be factored in.  One possibility may be to design courses that permit lighter weight
  forms of interaction.  For example, industry practitioners can be used as mentors on project based courses, meeting
  students for fortnightly review of progress.  Developing a range of engagement mechanisms that are properly advertised
  with indicative time commitments can also create a ``menu'' of options for practitioners to select from according to
  their availability.
\item The design and delivery of new higher education courses requires a considerable investment by the respective
  parties. In order to attract new organisations into this domain, we would recommend that the benefits to the
  organisation are stressed.  As we found, these arise in several ways.  Most immediately, the practitioners involved
  benefit from career enhancement, through the development and practice of new skills, and reflection on their own
  practices.  However, the wider organisation also benefits, through the internal reuse of materials, for example, and
  the opportunity for practitioners to advocate for process improvement based on these materials when they return to the
  organisation.
\item Consideration should be given as to the formality adopted for agreements between the collaborating
  academic institution and organisation.  In our case, an informal agreement was adopted between the coauthors. This was
  partly due to the good will that already existed mitigating the need for anything more formal.  Whilst this can
  provide greater flexibility and allow changes to happen more quickly, it also means that individual practitioners may
  find it more difficult to gain legitimacy for their involvement in the activity.  This can also mean that
  opportunities for participation may not be evenly distributed across the organisation, since it will be dependent on
  individual line managers recognising the value to the wider organisation.  A more formal agreement would mean that the
  activity has a recognised `owner' within the organisation, easing requests for resources etc.
\item Similarly, from the institutional perspective, steps should be taken to properly recognise and support the
  contribution of industry practitioners.  At the \UniversityofGlasgow, all the practitioners are registered with either
  affiliate or honorary staff status, depending on their involvement in the course.  This gives the
  practitioners academic profiles and also access to institutional resources, such as email and library subscriptions.
  However, the intermittent nature of their participation means that they may require additional support in navigating
  institutional processes, for example in recovering lost passwords, or ensuring that accounts are not disabled due to
  inactivity.  Other opportunities to participate in institutional activities should also be offered to assist with
  familiarisation.  For example, practitioners could be invited to participate in teaching ``Away Days'', that are used
  by faculty to reflect on strategic issues regarding curricula.  This also provides an opportunity for practitioners to
  build wider relationships with the faculty beyond the immediate course team.

\end{itemize}

In the future we plan to reflect further on our findings and recommendations and explore ways that practitioners can be
better supported in the co-design and delivery of higher education courses.  In addition, it would be desirable to repeat the retrospective exercise we conducted with other organisations and individuals that have participated in co-design with the \UniversityofGlasgow, and the other academic institutions that have delivered courses with \JPMorgan. This will help validate our findings to date and may reveal further recommendations that can widen the scope for collaboration. 

We believe that the overall collaboration
between \UniversityofGlasgow~ and \JPMorgan~ has been very successful and we are aware that versions of our courses have
now been adopted at two other higher education institutions, working with different practitioners within \JPMorgan.  As
part of our future work we will also seek the perspectives of practitioners involved in these courses, as well as
practitioners who have co-designed other courses at the \UniversityofGlasgow.

\begin{acks}
  The coauthors would like to thank our respective institutions, \UniversityofGlasgow~ and \JPMorgan.  We would also
  like to thank the Royal Academy of Engineering's Visiting Professorship scheme for supporting some of the
  practitioners in course development.

\end{acks}

\bibliographystyle{ACM-Reference-Format}
\bibliography{references}

\end{document}